\documentclass[10pt, conference, letterpaper, twocolumn]{IEEEtran}
\usepackage{amsmath,amssymb,amsfonts}
\usepackage{graphicx}
\usepackage{textcomp}
\usepackage[dvipsnames]{xcolor}

\usepackage{mathtools}  
\usepackage{amsmath}
\usepackage{amssymb}
\usepackage{tabulary}
\usepackage{booktabs}
\usepackage[inkscapearea=page]{svg}

\usepackage{paralist}

\usepackage{bbding}
\usepackage{pifont}
\usepackage{wasysym}
\usepackage{amssymb}
\usepackage{diagbox}

\usepackage[top=0.70in, left=0.58in, bottom=.96in, right=0.57in]{geometry}

\usepackage[font=small]{caption}
\usepackage{subcaption}
\usepackage{booktabs}
\usepackage{bbm}
\usepackage{algorithmic}
\usepackage{algorithm}
\usepackage[numbers, compress]{natbib}

\usepackage[hyphens]{url}

\usepackage{mathptmx}
\usepackage{amsmath}
\usepackage{amssymb}
\usepackage{amsthm}

\theoremstyle{definition}

\DeclareMathOperator*{\argmin}{arg\,min}
\newcommand\indicator{\mathbbm{1}}

\newcommand{\ie}{{\textit{i.e., }}}
\newcommand{\eg}{{\textit{e.g., }}}

\begin{document}

\title{
    % \LARGE{
    \huge{
    Multi-Task Decision-Making for Multi-User 360$^\circ$\\ Video Processing over Wireless Networks
    % Joint Communication and Computation Resource Allocation for Emerging mmWave Multi-User 3D Video Streaming Systems
    % }
    }
}
% {
\author{
    \IEEEauthorblockN{Babak Badnava\IEEEauthorrefmark{1}, Jacob Chakareski\IEEEauthorrefmark{2}, Morteza Hashemi\IEEEauthorrefmark{1}}
    \IEEEauthorblockA{
    \IEEEauthorrefmark{1} Department of Electrical Engineering and Computer Science, University of Kansas \\
    \IEEEauthorrefmark{2} College of Computing, New Jersey Institute of Technology
    % Universities Space Research Association (USRA)
    } 
    Emails: \{babak.badnava,mhashemi\}@ku.edu, jacobcha@njit.edu
}
% }
\maketitle
% \thispagestyle{plain}
% \pagestyle{plain}

% \setstretch{.95}
% \setlength{\textfloatsep}{0pt}

\begin{abstract}
We study a multi-task decision-making problem for $360^\circ$ video processing in a wireless multi-user virtual reality (VR) system
% We consider a multi-user joint rate adaptation and computation distribution problem in a millimeter wave (mmWave) virtual reality (VR) system.
that includes an edge computing unit (ECU) to deliver $360^\circ$ videos to VR users and offer computing assistance for decoding/rendering of video frames. However, this comes at the expense of increased data volume and required bandwidth.
% The VR system comprises an edge computing unit (ECU) that serves $360^\circ$ videos to VR users. In this system, VR users are assisted with the ECU that provides additional computational resources that can be used to process video frames (\ie decoding/rendering), at the expense of increased data volume and required bandwidth.
% \jacob{Can we put a little more meat here? for instance, we can describe what our framework comprises and talk about its components briefly. Do we have some contributions that we can cite here?}
%, with different spatio-temporal characteristics.
%The $360^\circ$ video streaming applications, compared to traditional 2D video streaming applications, require higher computational resources due to the additional complexities involved in encoding, decoding, spatial processing, stitching, and rendering.
% The $360^\circ$ videos need to be decoded and rendered before users could play them on their VR headset.
% , need to be decoded and rendered, which can take place at an edge computing unit (ECU) or VR headsets, before users could play them on their VR headset.
To balance this trade-off, we formulate a constrained quality of experience (QoE) maximization problem in which the rebuffering time and quality variation between video frames are bounded by user and video requirements.
% In this system, VR users are assisted with the ECU to process (\ie decode/render) $360^\circ$ videos.
% The ECU provides additional computational resources that can be used for processing video frames, at the expense of increased data volume and required bandwidth.
%. This processing, however, significantly increases the data sizes and  required bandwidth. 
% to enhance the users' QoE, however, introduces a higher bandwidth requirement, which in certain network conditions may not be available.
To solve the formulated multi-user QoE maximization, we leverage
% the state-of-the-art 
deep reinforcement learning (DRL) for \underline{m}ulti-\underline{t}ask 
% develop a sample efficient 
\underline{r}ate adaptation and \underline{c}omputation distribution (MTRC).
% \jacob{Why is it that we decided to use deep learning to begin with? Do we explain that here?}
% DRL methods make decisions exclusively by analyzing the outcomes of previous decisions and adapting accordingly to improve the users' QoE.
The proposed MTRC approach 
% The proposed MTRC algorithm 
does not rely on any predefined assumption about the environment and relies on video playback statistics (\ie past throughput, decoding time, transmission time, etc.), video information, and the resulting performance to adjust the video bitrate and computation distribution.
We train MTRC with real-world wireless network traces and $360^\circ$ video datasets to obtain evaluation results in terms of the average QoE, peak signal-to-noise ratio (PSNR), rebuffering time, and quality variation. 
% a comprehensive evaluation in terms of 
% various factors involved in users' QoE.
% various network scenarios and resource allocation mechanism at ECU.
Our results indicate that the MTRC improves the users' QoE compared to state-of-the-art rate adaptation algorithm.
% perceived video quality by $3.08$ dB to $4.49$ dB compared to the state-of-the-art rate adaptation algorithm. 
% Specifically, we show a $3.08$ dB to $4.49$ dB improvement in video quality in terms of PSNR, a $12.5$X to $14$X reduction in rebuffering time, and a $3.07$ dB to $3.96$ dB improvement in quality variation.
Specifically, we show a $5.97$ dB to $6.44$ dB improvement in PSNR, a $1.66$X to $4.23$X improvement in rebuffering time, and a $4.21$ dB to $4.35$ dB improvement in quality variation.
% \jacob{Can we include some further results and briefly outline them here?}
% baselines and provides a close-to-optimal performance.
% Experimental results indicate that MTRC improves QoE per group of picture (GoP) on average by $1.9\%$ compared with ECU proportional resource allocation.
\end{abstract}

\begin{IEEEkeywords}
Quality of experience, wireless networks, $360^\circ$ video processing, edge computing, mobile VR systems.
\end{IEEEkeywords}

% ###########################################
\vspace{-.05in}
\section{Introduction}\label{sec:Intro}
\vspace{-.05in}
Next-generation wireless networks (6G and beyond) will enable new use cases and applications that demand significantly higher computational power and bandwidth. Augmented reality (AR), virtual reality (VR), and extended reality (XR) are examples of such applications~\cite{SeaGate-2019-State,Chakareski-2022-Towards,Chakareski-2023-Millimeter,statistica-2024-vr,chakareski2019uav,ChakareskiK:23}.
For instance, VR applications capture entire spherical scenes and stream high-fidelity 360° video content to create an immersive experience for users. This process demands substantial computational and communication resources. Unlike traditional 2D video streaming, $360^\circ$ video streaming requires additional computational power for encoding, decoding, spatial processing, stitching, and rendering~\cite{Chakareski-2020-6DOF,Chakareski:18}.
% In particular, in contrast to traditional 2D video streaming, $360^\circ$ video streaming  requires computational resources 
% for encoding, decoding, spatial processing, stitching, and rendering~\cite{Chakareski-2020-6DOF}.
In fact, $360^\circ$ video decoding entails 
spatio-temporal transformations for spherical projection.
Viewport-adaptive streaming further increases the computational complexity by dynamically adjusting video segments based on the viewer's field of view (FoV). 
Moreover, $360^\circ$ videos typically have higher resolution and larger file sizes compared to 2D videos, leading to higher bandwidth requirements. 
To satisfy these requirements,  multi-access edge computing (MEC) and high-bandwidth mmWave wireless networks have been proposed in prior works~\cite{Chakareski-2020-6DOF, Hsu-2020-MEC, Gupta-2023-mmWave, Ren-2019-Edge}.

% Although an extensive amount of work has been focused on maximizing QoE for 2D video streaming
Numerous studies have focused on maximization of QoE for 2D video streaming (see, for example, ~\cite{Mao-2017-Neural, Spiteri-2020-BOLA,Badnava-2022-qoe,  chakareski2008distributed, reis2010distortion, ChakareskiATWG:04,ChakareskiAWTG:04c}). On the other hand, due to the increased computational and communication requirements for $360^\circ$ video streaming, it is necessary to develop  joint resource allocation to ensure a satisfactory QoE for VR users. Such a joint optimization problem should incorporate several factors including (i) available communication data rates provided by the underlying wireless network, (ii) computational resources provided by the VR headset and/or MEC unit, and (iii) {spatio-temporal characteristics} of $360^\circ$ videos.

\begin{figure}
    \centering
    \vspace{-.5cm}
    \includegraphics[width=.9\linewidth, trim={4.3cm 4.1cm 0cm 0cm},clip]{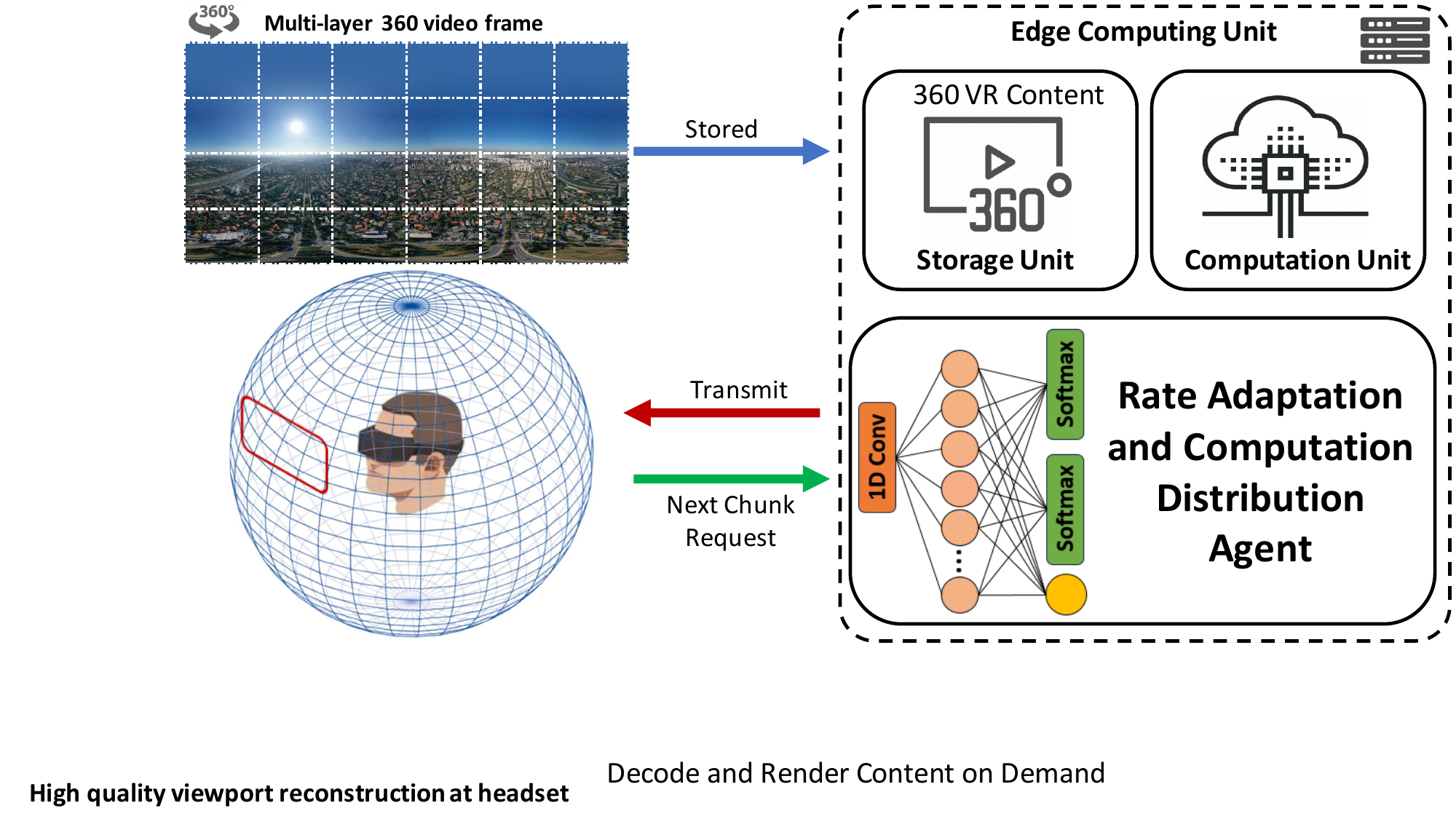}
    \caption{Edge-assisted VR system model: multiple VR headset connected to an edge computing unit through a mmWave network}
    \label{fig:system-model}
    % \vspace{-.05in}
\end{figure}

% 
% Integration of edge computing and dual connectivity wireless technologies have been investigated in by \cite{Chakareski-2020-6DOF}.
% A multitude of prior studies have investigated various aspects of VR systems.
Extensive amounts of prior work have focused on different aspects of VR systems. For example, the authors in~\cite{Chakareski-2020-6DOF} propose a dual connectivity streaming system in which mmWave and Wi-Fi links are integrated to enable six degrees of freedom VR-based remote scene immersion. Furthermore, the authors in 
\cite{Hou-2021-Predictive} propose a FoV prediction algorithm so that the predicted view is encoded in relatively high quality and transmitted in advance, which reduces the latency.
% to alleviate the network-imposed constraints by reducing the bandwidth requirement, 
\citeauthor{Hsu-2020-MEC} in \cite{Hsu-2020-MEC} investigates the optimization of caching and computing at the edge server to improve the QoE of users.
\citeauthor{Gupta-2023-mmWave} in \cite{Gupta-2023-mmWave} leverage an edge server for decoding to maximize the smallest immersion fidelity for the delivered $360^\circ$ content across all VR users. 
The authors in \cite{Ren-2019-Edge}
present an MEC computing framework to optimize energy efficiency and processing delay
for AR applications. 
Moreover, \cite{Chakareski-2021-Full} investigates the rate-distortion characteristics of ultra-high definition (UHD) $360^\circ$ videos.
% to alleviate videos-imposed constraints.
% This is the closest work to our work, however, we investigate computation decisions in more details by distinguishing between decoding and rendering decisions.
% However, to the best of our knowledge 

Despite extensive research on VR systems, to the best of our knowledge, no prior work has considered multi-task video processing for multi-user wireless VR systems.
% fully integrated all the previously mentioned constraints for 
% and there are gaps in developing dynamic decision-making algorithms that maximize multi-user QoE while incorporating communication, computation, and video constraints.
% However, to the best of our knowledge, no prior research fully integrates all the previously mentioned constraints for multi-user VR systems that use mmWave networks.  
% arena under mmWave network.
Therefore, in this paper we formulate a multi-task edge-assisted video quality maximization framework for $360^\circ$ video streaming on wireless networks as depicted in Fig. \ref{fig:system-model}.
% In this framework, the computation required for $360^\circ$ videos can be performed by an ECU, located in the proximity of the VR arena, or by users' headsets.
In this framework, the computational tasks (\ie decoding and rendering) necessary for processing $360^\circ$ videos  may be executed either by an ECU or by the users' headsets themselves.
% In this framework, $360^\circ$ videos can be decoded and/or rendered by an ECU, located in the proximity of the VR arena, or by users' headsets.
On one hand, the ECU has more computational resources to process the $360^\circ$ videos faster, which leads to lower computational latency and higher QoE for users.
However, decoding and rendering by the ECU introduces a higher bandwidth requirement since processed videos have much larger sizes. Therefore, this leads to higher communication latency, thus degrading the QoE.
Furthermore, the ECU provides its best performance for a certain number of users due to limited computational resources.

To capture these trade-offs and jointly optimize communication and computation resource allocation, we present a novel learning-based decision-making algorithm, called MTRC, which considers the interaction between the communication and computation requirements of $360^\circ$ videos.
Learning-based methods, particularly DRL, do not depend on predefined models or system assumptions \cite{Badnava-2023-Energy, Badnava-2021-Spectrum}. 
Rather, they learn to make decisions exclusively by analyzing the outcomes of previous decisions and adapting accordingly.
% which makes them suitable for addressing the challenges involved in such a complex environment.
% \jacob{We do not explain why we chose to use ML to solve the problem. Why not a traditional algorithm?}
MTRC leverages a state-of-the-art DRL method to learn the optimal computation distribution (\ie ECU or headset) and the video bitrate in the VR arena by considering the playback statistics and video information.
In summary, the main contributions of this paper are as follows:
\begin{itemize}
    \item We consider an edge-assisted wireless VR streaming system in which an ECU provides $360^\circ$ videos to VR users.  
    We formulate a constrained video quality maximization problem, in which rebuffering time and video quality variation are bounded by user and video requirements, to find the best policy w.r.t network condition and spatio-temporal characteristics of multi-layer $360^\circ$ videos.
    \item We develop a multi-task learning-based algorithm to find the optimal computation distribution and video bitrate for VR users to maximize their QoE. Our MTRC agent observes the playback statistics (\ie past throughput, decoding/transmission time, etc.) and video information, then decides the optimal bitrate and computation distribution to decode/render the $360^\circ$ video.
    \item We develop a $360^\circ$ VR streaming simulator using a real $360^\circ$ video dataset, real-world VR user navigation information, and real-world mmWave network traces. 
    Then, we perform an extensive simulation to analyze the behavior of our proposed method.
    We show that the proposed MTRC algorithm improves the PSNR by $5.97$ dB to $6.44$ dB, rebuffering time by $1.66$X to $4.23$X, and quality variation by $4.21$ dB to $4.35$ dB.
\end{itemize}
% The rest of this paper is organized as follows. 
% In Section \ref{sec:system}, we present the VR system model.
% In Section \ref{sec:problem}, we formulate the multi-user QoE maximization problem followed by our solution in Section \ref{sec:solution}.
% % to the multi-user QoE maximization problem, called MTRC. 
% In Section \ref{sec:evaluation}, we provide the simulation results, and Section \ref{sec:conclusion} concludes the paper.

% \jacob{We can save space in the end. For instance, an outline of the rest of the paper is not really needed. The contributions of the paper highlighted above can be shortened too.}

\section{System Model}\label{sec:system}
% \vspace{-.05in}
We consider a multi-user $360^\circ$ VR video streaming application 
with $N$ users, each connected to the ECU via a wireless access point.
As depicted in Fig. \ref{fig:system-model}, the users are equipped with VR headsets and request $360^\circ$ videos that are stored on the ECU.
The videos need to be processed (\ie decoded/rendered) and transmitted to VR headset before users can play them.
Upon receiving a request for a video segment, a decision-making agent (\ie MTRC agent) makes a joint decision on the video bitrate allocated to each user and the computation distribution (\ie decoding and rendering on the ECU or on the headset).
Then, computational resources will be allocated to process each video segment. 

The ECU and all VR headsets are equipped with computational resources (\ie CPU and GPU) to process the video segments.
If the decision-making agent decides to process the video segment on the ECU, the ECU's computational resources are shared among users to process the video segments and then transmit the video segments to the users.
However, if the decision-making agent decides to process the video segment on users' headset, the ECU sends the video segments to users and the preparation takes place at the headsets. 
Next, we present the models on multi-layer $360^\circ$ videos,  user headsets, and ECU.

\begin{figure}
    \centering
    \includegraphics[width=.95\linewidth, trim={4.3cm 5.2cm 4.3cm 3.8cm},clip]{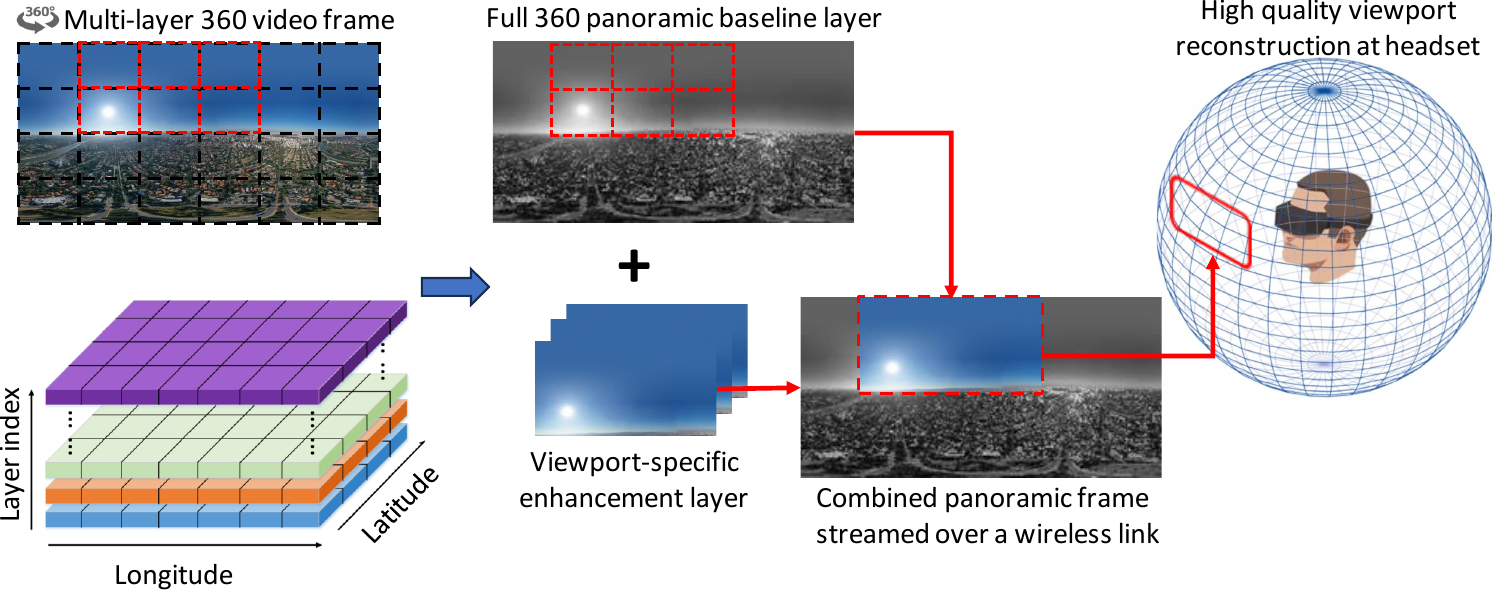}
    \caption{
    % Multi-layer $360^\circ$ video model: 
    Viewport-specific enhancement layers and a baseline layer are transmitted to a VR headset via a wireless link to be reconstructed.}
    \label{fig:video-model}
    % \vspace{-.05in}
\end{figure}
\textbf{Multi-layer $360^\circ$ Video Model:}
We consider
% a set of $360^\circ$ videos stored on ECU and leverage 
the scalable multi-layer $360^\circ$ video viewpoint tiling design~\cite{Chakareski-2020-6DOF}.
As depicted in Fig. \ref{fig:video-model}, each panoramic $360^\circ$ video frame is partitioned into $\mathcal{L}$ tiles arranged in a $L_H \times L_V$ grid.
A block of consecutive video frames, compressed together with no reference to others, creates a group of pictures (GoP) or video segment.
Each video is divided into $M$ GoP with fixed time duration of $\Delta t$, and $L$ layers of increased immersion fidelity for each tile in a GoP exist.
The first layer is called the base layer, and the remaining layers are denoted as enhancement layers.

Each enhancement layer increases the video bitrate, hence the video quality.
% Next, we use video bitrate information, user's viewport information, and compression reduction factor to calculate the size of a compressed GoP, decoded GoP, and rendered GoP.
We denote $e^m_n \in \left\lbrace 0,1 \right\rbrace^{L}$ as a one-hot vector that determines the number of enhancement layers included in the $m^{th}$ GoP requested by the $n^{th}$ user.
% and denote $w^m_n \in \left\lbrace 0, 1 \right\rbrace^{L_H \times L_V}$ as user's viewport information, where $w^m_n \left[ i, j\right] = 1$ when tile with index $(i, j)$ is in user's viewport.
% Note that each video quality index corresponds to an enhancement layer, which means as we increase the number of enhancement layers, we increase the video quality.
Then, the size of the $m^{th}$ compressed GoP tile, denoted by $d(e^m_n)$, is determined by summing over all tiles' bitrates in the user's viewport.
We assume a positive compression reduction factor of $\beta < 1 $, which leads to $d(e^m_n)/\beta$ for the size of the $m^{th}$ decoded GoP tile.
After decoding, GoPs need to be rendered as well, which leads to an increase in size by a factor of $\alpha \geq 2$.
Hence, the size of the $m^{th}$ GoP after decoding and rendering is determined by $\alpha d(e^m_n)/\beta$.
% Next, we will model the edge server and user's VR headset.

% \vspace{-.05in}
% \subsection{User Headset Model}
% \vspace{-.05in}
% The VR headset is connected to the ECU via a mmWave wireless network. 
% The ECU transmit videos to VR headsets through this wireless link.
% The ECU chooses to send the raw video (\ie compressed video), decoded video, or decoded and rendered video to each of the headsets.

% Next, we model the buffer dynamics.

% $$
% \bar{Z}_{rend.} = \left[ .... \right]_{1\times N} \\ 
% \bar{Z}_{dec.} = \left[ .... \right]_{1\times N}
% $$

\begin{figure}
    \centering
    \includegraphics[width=.8\linewidth, trim={0cm 0.4cm 0cm 0cm},clip]{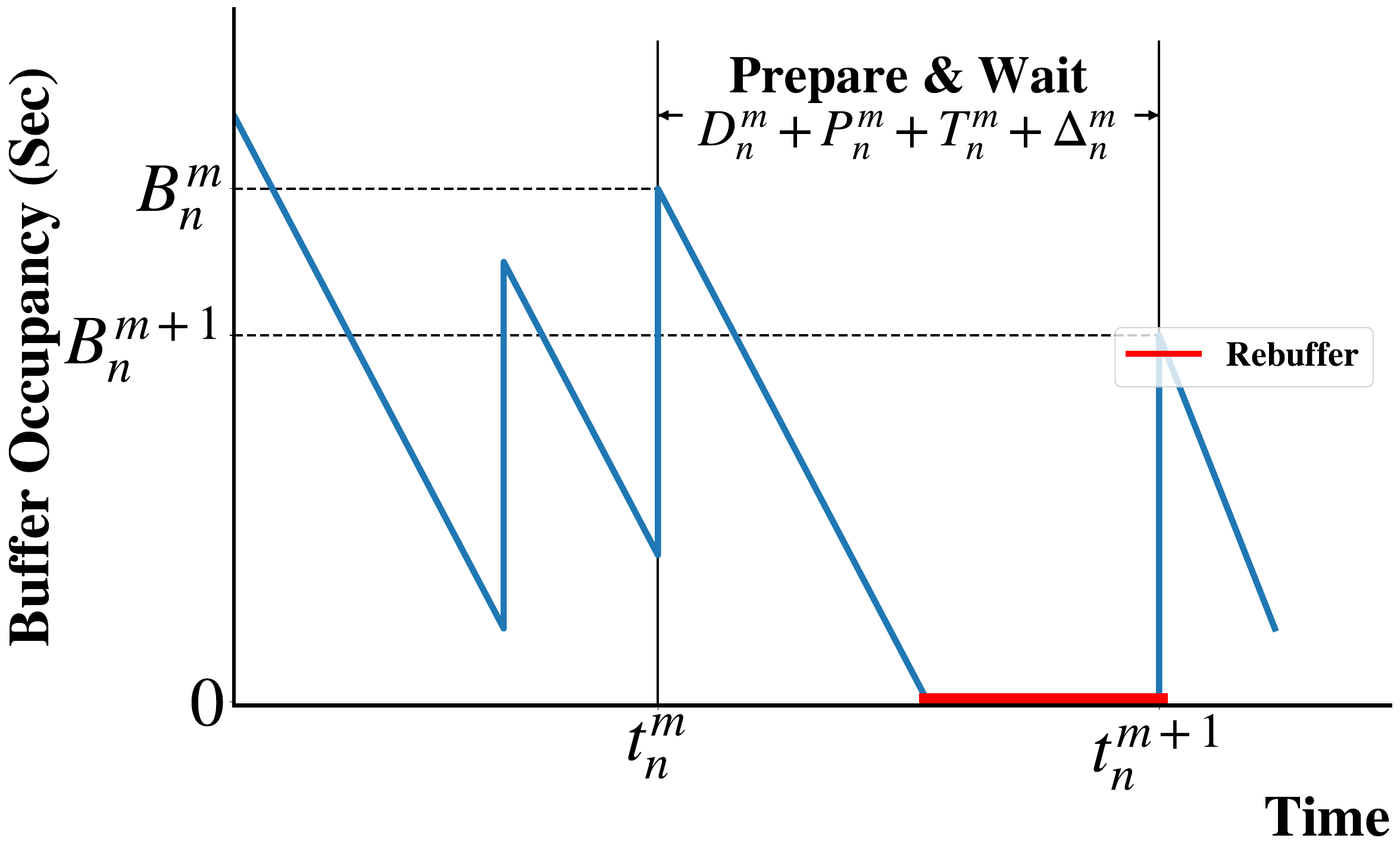}
    \vspace{-.4cm}
    \caption{VR headset buffer dynamics}
    \label{fig:buffer-dynamics}
    % \vspace{-.25in}
\end{figure}

\textbf{Video Retrieval and Buffering:}
The VR headsets retrieve and store rendered videos in a buffer with a fixed duration time.
Fig. \ref{fig:buffer-dynamics} illustrates the buffer dynamics of the $360^\circ$ video streaming application.
The GoPs need to be processed in order to be buffered on headsets.
At time $t^m_n$, the $n^{th}$ user requests the $m^{th}$ GoP.
Then, the GoP will be processed (\ie either on the ECU or the headset) and buffered on headsets.
% The preparation phase includes decoding, rendering, and transmission of GoP.
% The decoding time, denoted by $D^m_n$, and rendering time, denoted by $P^m_n$, depend on the amount of computational resources allocated to that GoP.
% which varies depending on where the decoding/rendering process take place (\ie either on the ECU or on the VR headset). 
% Later, we will show how decoding/rendering times are modeled both on the ECU and on VR headsets.
The preparation of $m^{th}$ GoP involves three key stages: the decoding time $D^m_n$, rendering time $P^m_n$, and transmission time $T^m_n$. Once the GoP is processed and buffered, the user waits for $\Delta^m_n$ seconds until requesting the next GoP. Thus, the next request time is:
% This waiting time is small enough not to cause any rebuffering events.
$
% \small
% \begin{equation*}
% \small
% \begin{aligned}
    t^{m+1}_n = t^m_n + D^m_n + P^m_n + T^m_n + \Delta^m_n.
% \end{aligned}
% \end{equation*}
$
% \normalsize
The buffer occupancy evolves as GoPs are being prepared, and the video is being played by the user.
The buffer occupancy of user $n$ increases by $\Delta t$ seconds after receiving GoP $m$.
Let $B^m_n = B_n(t^m_n)$ denote the buffer occupancy of the $n^{th}$ user at $t^m_n$.
% when the headset requests the $m^{th}$ GoP. 
Then, we have: 
% $$
% \small
\begin{equation*}
% \small
\begin{aligned}
    B^{m+1}_n = B_n (t^{m+1}_n)
    = 
    \left(
        \left(
        B^m_n - P^m_n - D^m_n - T^m_n
        \right)_+
        + \Delta t - \Delta^m_n
    \right)_+.
\end{aligned}
\end{equation*}
% $$
% \normalsize
Here, the notation $\left(x\right)_+ = \max \left\lbrace 0, x \right\rbrace$ ensures that the buffer occupancy is non-negative.
If the process and transmission times take longer than the amount of GoP stored in the buffer (\ie $B^m_n < P^m_n + D^m_n + T^m_n$), then \textit{rebuffering} happens as shown in Fig. \ref{fig:buffer-dynamics}.
We also assume that the waiting time $\Delta^m_n$ is zero, except when the buffer is full, which the headset waits until the buffer has enough space to accommodate the next GoP, which leads to:
% which means that the next GoP will be requested immediately after complete reception of the current GoP.
% The one exception is when the buffer is full, then the headset waits for until the buffer have enough space to accommodate the next GoP, which leads to:
% \small
$
% \begin{equation*}
% \small
%     \begin{aligned}
        \Delta^m_n = \left( 
            \left( B^m_n - P^m_n - D^m_n - T^m_n \right)_+ 
            + \Delta t - B^{max}_n
        \right)_+.
%     \end{aligned}
% \end{equation*}
$

% \normalsize
% Next, we will model the decoding and rendering computational requirements on the VR device.

\textbf{VR Headset Decoding and Rendering Model:}
The VR headsets are equipped with a CPU and GPU to process the videos, and
each VR headset provides a maximum decoding speed of $\bar{Z}^{dec.}_{n}$, a maximum rendering speed of $\bar{Z}^{rend.}_{n}$, where $n$ is the index of the VR headset.
We incorporate a decoding and rendering model developed by \cite{Chakareski-2023-Millimeter} to compute the decoding and rendering time for each GoP.
In this model, the decoding time of the $m^{th}$ GoP for the $n^{th}$ user is assessed as $\Tilde{si}(e^m_n)/ \bar{Z}_{dec.}^n$, where $\Tilde{si}(.)$ returns the computational complexity of decoding a GoP (in bits), which is the induced data rate associated with the current viewport \ie $si(e^m_n) = d(e^m_n)$.
% given the video bitrate and the user's viewport information.
Similarly, the rendering time is modeled as $si(e^m_n)/ \bar{Z}^{rend.}_n$, where $si(.)$ returns the rendering computational complexity of a GoP (in bits), which is the induced data rate after decoding of the GoP $si(e^m_n) = d(e^m_n) / \beta$.
Note that we assume that the viewport information is available on the headset.

\textbf{Edge Computing Unit (ECU) Model:}
The ECU provides additional computational resources to assist VR users with decoding and rendering.
This additional computational resources provide a maximum decoding speed of $Z^{dec.}_{ECU}$ and a maximum rendering speed of $Z^{rend.}_{ECU}$, which is shared among the users that decode/render their GoP on the ECU.
We assume that the decision-making time is negligible since the decoding and rendering tasks are dominant overheads.
Then, we incorporate a similar computation model on the ECU.
% The ECU stores all the $360^\circ$ videos.
% Videos need to be processed before streaming to users. 
% The video processing is composed of decoding and rendering of $360^\circ$ videos.
% As mentioned earlier, decoding and rendering can take place at both edge server and VR headset, which is decided by the edge server.
% The ECU is equipped with a GPU, which provides a maximum decoding speed of $Z^{dec.}_{ECU}$ and a maximum rendering speed of $Z^{rend.}_{ECU}$.
% to process high-fidelity $360^\circ$ videos before streaming them to users through a mmWave wireless network.
% The ECU is equipped with a GPU to process high-fidelity $360^\circ$ videos before streaming them to users through a mmWave wireless network.
% The ECU provides a maximum decoding speed of $Z^{dec.}_{ECU}$, a maximum rendering speed of $Z^{rend.}_{ECU}$.
% ECU's computational resources (\ie decoding and/or rendering resources) are shared among users to provide extra computational power for users to decode/render their videos.
% We continue this section by modeling the computational and communication requirements of VR system model.

% \textbf{Video Decoding and Rendering Model:}
The decoding starts immediately after receiving the request for the next GoP if the decision-maker decides to decode the GoP on the ECU. This leads to
% The ECU distributes its decoding computational resources among those users that it decided to decode their GoPs on the ECU.
% Each video needs to be decoded first and then rendered from 2D to 3D FoV.
% We assume that the decision-making time is negligible, since the decoding and rendering tasks are dominant overheads. 
% Thus, in case of decoding at the ECU, the decoding starts immediately after receiving the request for next GoP, which leads to
% Then, the ECU starts allocating resources to decode the requested GoP, which leads us to
% \begin{equation}
    % \begin{aligned}
$
        % \frac{
            \Tilde{si}(e^m_n) /
        % }{
            % \sum_{s=t^m_n}^{t^{m+1}_n - T^m_n - P^m_n - \Delta^m_n} \Psi^m_n
            \Psi^m_n
        % }
$
    % \end{aligned}
% \end{equation}
seconds of decoding time.
% Here, the function $\Tilde{si}(.)$ returns the decoding computational complexity of a GoP.
% However, since the ECU does not have access to the actual user's viewport information, it needs to estimate the viewport information.
% Hence, $\Tilde{w}^m_n$ denotes the estimated viewport information of the $n^{th}$ user while watching the $m^{th}$ GoP.
Here, $\Psi^m_n$ denotes the amount of decoding resources, out of $Z^{dec.}_{ECU}$, allocated to the $n^{th}$ user to decode the $m^{th}$ GoP.
Similarly, the ECU rendering time is modeled as $si(e^m_n) / \Theta^m_n$, where $\Theta^m_n$ denotes the amount of rendering resources, out of $Z^{rend.}_{ECU}$, allocated to the $n^{th}$ user for the $m^{th}$ GoP.
Note that the total amount of computational resources allocated to users cannot exceed the maximum available resources, which means that $\sum_{n=1}^N \Psi^m_n \leq Z^{dec.}_{ECU}$ and $\sum_{n=1}^N \Theta^m_n \leq Z^{rend.}_{ECU}$ must be satisfied for each GoP.

\textbf{Wireless Links Model:}
The ECU transmits GoPs through a mmWave wireless network.
% The mmWave networks enjoy high bandwidth but suffer from higher path loss due to free space, oxygen absorption, and/or penetration losses.
% The mmWave wireless technologies operate at mmWave bands, where more bandwidth is available, however suffer from higher path loss due to free space, oxygen absorption, and/or penetration losses.
% This loss is compensated by employing directional beamforming techniques.
% It also provides multi-user connectivity through directional beamforming.
% Benefiting from both larger bandwidth and beamforming techniques, mmWave enables multi-Gbps transmission rates.
The expected transmission rate for a GoP is modeled as
$
% \begin{equation}
% \begin{aligned}
C^m_n = \frac{1}{t_e - t_s} \int_{t_s}^{t_e} C_s^n ds
% = \frac{1}{t_e - t_s} \sum_{s=t_s}^{t_e} C_s^n
% \end{aligned}
% \end{equation}
$
where, $t_s$ and $t_e$ are transmission start and end times, respectively, and $C^s_n$ is the throughput provided by the wireless channel for the $n^{th}$ user.
Hence, the transmission time for a compressed GoP follows
$
% \frac{
    d(e^m_n) /
% }{
    C^m_n
% }
.
$
Similarly, we can model the transmission time for decoded and rendered GoPs. 

% Hence, the transmission time for a compressed GoP, which will be started immediately after reception of GoP request, will be: 
% $$
% \frac{
%     d(r^m_n, w^m_n)
% }{
%     C^m_n(t^m_n, t^{m+1}_n - D^m_n - P^m_n - \Delta^m_n)
% }.
% $$
% The transmission for a GoP that is decoded at the edge server starts immediately after finishing the decoding, which leads to transmission time of:
% $$
% \frac{
%     d(r^m_n, \Tilde{w}^m_n)/\beta
% }{
%     C^m_n(t^m_n + D^m_n, t^{m+1}_n - P^m_n - \Delta^m_n)
% }.
% $$
% Furthermore, for GoPs that are a decoded and rendered at the edge server, the transmission starts immediately after the completion of projection, which will lead to:
% $$
% \frac{
%     \alpha  d(r^m_n, \Tilde{w}^m_n)/\beta
% }{
%     C^m_n(t^m_n + D^m_n + P^m_n, t^{m+1}_n - \Delta^m_n)
% }.
% $$

\vspace{-.05in}
\section{Problem Formulation}\label{sec:problem}
\vspace{-.05in}
% We begin this section by defining the QoE for users of a $360^\circ$ video streaming application. 
% Then we formulate our multi-user QoE maximization problem and discuss the challenges involved in solving such a problem.
% We will finish this section by discussing the challenges that DRL based approaches need to tackle in order to solve this problem.
% In this section, we formulate our multi-user QoE maximization problem for a $360^\circ$ video streaming application and discuss the challenges involved in solving such a problem.

% \textbf{Quality of Experience (QoE):}
Given the presented system model, we aim to improve QoE for multi-user $360^\circ$ video streaming. To this end, we consider three major factors that impact QoE. The first factor is the Average Video Quality (AVQ) defined as the average per-GoP video quality for tiles in the user's FoV, expressed as: $Q(e_n)=\frac{1}{M} \sum_{m=1}^{M} q(e^m_n)$. 
While there are various choices for $q(.)$~\cite{Mao-2017-Neural}, 
% the function $q(.)$ maps the enhancement layer index to the quality perceived by the user.
we use PSNR for the viewer's FoV \cite{Yu-2015-Framework}, which can be calculated using the video distortion \cite{Chakareski-2021-Full} as $q(e^m_n) = 10 \log_{10}(255^2 / MSE^m_n)$, where $MSE^m_n$ is the distortion of the $m^{th}$ GoP. The distortion has an inverse relation with the video bitrate, which is determined by the number of enhancement layers streamed to the user~\cite{Chakareski-2021-Full}.

% in order to achieve higher long-term user engagement.
% While users may differ in their preference, there are some key contributing factors in QoE such as:
% \begin{enumerate}
%     % \item \textit{ :}  
%     % \begin{equation*}
%     % \small
%     %     \operatorname{\textrm{AVQ}}=\frac{1}{M} \sum_{m=1}^{M} w^m_n \odot q(e^m_n)
%     % \end{equation*}
%     % \item 
%     \item 
% \end{enumerate}

The second factor that impacts perceived QoE is the average quality variation (AQV) that captures quality variation in the user's FoV from one GoP to another. Therefore, we have $V(e_n) = \frac{1}{M-1} \sum_{m=1}^{M-1} 
    \left|q(e^{m+1}_n) - q(e^m_n)\right|$. The third factor that affects QoE is rebuffering. Rebuffering occurs if the process time of a GoP is larger than the buffer occupancy level when the GoP was requested. Thus, the total rebuffering time (RT)  is given by:
    $S(e_n, \phi_n) = \sum_{m=1}^{M} \left( D^m_n + P^m_n + T^m_n - B^m_n \right)_+$.
% \bb{
Several previous studies~\cite{Mao-2017-Neural, Spiteri-2020-BOLA, wang2023bones} have defined the $n^{th}$ user's QoE of GoP $1$ through $M$ by a weighted sum of the aforementioned components:
\begin{equation}
\label{eq:qoe}
% \small
\begin{aligned}
    \hat{QoE}^n_M = Q(e_n)   -  \mu_0 S(e_n, \phi_n) - \mu_1 V(e_n), 
\end{aligned}
\end{equation}
where, $\mu_0$ and $\mu_1$ are non-negative weighting parameters corresponding to user sensitivity to rebuffering time and quality variation, respectively.
Although this QoE metric allows us to model varying user preferences \cite{Mao-2017-Neural, Yin-2015-Control}, setting the values of $\mu_0$ and $\mu_1$ for various users' requirements is not straightforward. To resolve this issue, we will focus on optimizing the average video quality with constrained rebuffering time and average quality variations.  

\textbf{Multi-Task QoE Maximization Problem:}
% We are now ready 
To formulate the problem of multi-task QoE maximization, we define two sets of communication and computation decision variables.  
In particular, $\phi^m_n \in \left\{ 0, 1 \right\}^{3}$ is a binary vector of size three with one active element (\ie a one-hot vector), which determines where the decoding and rendering take place for each GoP and user.
There are three states for $\phi^m_n$ as follows: (i) $\phi_{m, 0}^n = 1$ corresponds to decoding and rendering on the ECU, (ii) $\phi_{m, 1}^n = 1$ if we decode on ECU, but render on headset, and (iii) $ \phi_{m, 2}^n = 1$ implies that both decoding and rendering happen on headset. 
 % \begin{equation}
 % \label{eq:phi-states}
 % \small
 %  % \phi_{m, i}^n = 1 : 
 %  \phi^m_n: 
 %  % \phi^m_n[i] :
 %  \begin{cases}
 %        \phi_{m, 0}^n = 1  \quad \Rightarrow \; \text{Decode \& Render on ECU,} \\
 %        \phi_{m, 1}^n = 1 \quad \Rightarrow \; \text{Decode on ECU \& Render on headset,}\\
 %        \phi_{m, 2}^n = 1 \quad \Rightarrow \; \text{Decode \& Render on headset.}
 %        % i = 0 \; \Rightarrow \; \text{Decode and Render on ECU,} \\
 %        % i = 1 \; \Rightarrow \; \text{Decode on ECU,}\\
 %        % i = 2 \; \Rightarrow \; \text{Decode and Render on headset.}
 %        \end{cases}
 % \end{equation}
% If $\phi^m_n\left[0\right] = 1$, both decoding and rendering are performed by the ECU, thus the ECU allocates computational resources for decoding and rendering of GoP $m$ requested by user $n$.
% If $\phi^m_n\left[1\right] = 1$,  the ECU performs only decoding, and rendering takes place at the VR headset of $n^{th}$ user.
% If $\phi^m_n\left[2\right] = 1$, both decoding and rendering are performed by the user's headset.
% For the sake of notation, we use $\phi^{m, i}_n$ instead of $\phi^m_n\left[i\right]$. 

In addition to the location of the computation, we consider the rate allocation decision variable $e^m \in \{0, 1\}^{N\times L}$ that determines how many enhancement layers should be streamed to each user in the VR arena. 
In addition to these decision variables, ECU computation resources (\ie decoding resources $\psi_n^m$  and rendering resources $\theta_n^m$) should be allocated to the users,
% who are determined to decode and/or render on the ECU
which we assume that they have been allocated proportional to user's requirements. therefore, we formulate the following optimization problem:
% outlined in Eq. \ref{eq:qoe-maximization}:
\begin{figure}[ht]
\hrulefill
\small
% \begin{equation}
% \label{eq:qoe-maximization}
% \resizebox{\hsize}{!}{
\begin{subequations}\label{eq:qoe-maximization}
\begin{alignat}{5}
    \max_{\mathbf{\phi_n, e_n}} \quad\;  Q(\mathbf{e_n}) & \tag{\ref{eq:qoe-maximization}} \\
    \text{s.t.}
    \quad  S(\mathbf{e^n}, \mathbf{\phi^n}) & \leq \mathcal{H}_0 \label{eq:rt-const} \\
    \quad  V(\mathbf{e^n})  & \leq \mathcal{H}_1 \label{eq:qv-const} \\
    % \quad & \sum_{n=1}^N \zeta_s^n \leq K \quad \forall s \\
    % \quad & \min \frac{1}{S} \sum_{s=1}^S \norm{\zeta_{s+1} - \zeta_s} \quad \forall s \\
    % \quad & \min \norm{w - \hat{w}}{} \quad \forall m, n \\
    \quad B^{m+1}_n &= 
    \left(
        \left(
        B^m_n - P^m_n - D^m_n - T^m_n
        \right)_+
        + \Delta t - \Delta^m_n
    \right)_+, \; \forall m \label{eq:buffert-b}\\
    \quad   t^{m+1}_n &= t^m_n \nonumber\\
    \quad &+ \phi_{m,1}^n \left[ 
    \frac{
            \Tilde{si}(e^m_n)
        }{
            \Psi^m_n
        } + \frac{
            si(e^m_n)
        }{
            \Theta^m_n.
        } + \frac{
             \alpha  d(e^m_n)/\beta 
        }{
            C^m_n
        } \right]
    \nonumber \\ &+ \phi_{m,2}^n \left[
        \frac{
            \Tilde{si}(e^m_n)
        }{
            \Psi^m_n
        } + 
        \frac{
            si(e^m_n)
        }{
            Z^{rend.}_{n}
        } + 
        \frac{
            d(e^m_n)/\beta
        }{
            C^m_n
        } \right] \nonumber \\
    \quad &+ \phi_{m,3}^n \left[
        \frac{
            \Tilde{si}(e^m_n)
        }{
            Z^{dec.}_{n}
        }
    +   \frac{
            si(e^m_n)
        }{
            Z^{rend.}_{n}
        } + 
        \frac{
            d(e^m_n)
        }{
            C^m_n
        }
    \right] 
    + \Delta^m_n,  \; \forall m  \label{eq:buffert-t} \\
    \quad \sum_{n=1}^N  \psi^m_n &\leq Z^{dec.}_{ECU} \quad \forall m \quad, 
    \quad \sum_{n=1}^N \theta^m_n \leq Z^{rend.}_{ECU} \quad \forall m \label{eq:comp-const}
    % \quad \sum_{i=1}^{3} \phi_{i,m}^n &= 1 \quad \forall m,n \\
\end{alignat}
% }
% \end{equation}
\end{subequations}
\hrulefill
% \vspace{-.1in}
\end{figure}

There are several key challenges that need to be addressed before solving the optimization problem outlined in Eq. \ref{eq:qoe-maximization}.
This is a constrained optimization problem, in which we maximize the average video quality such that the rebuffering time and average quality variation are constrained to a bound.
% In general, multi-objective optimization problems have a set of so-called Pareto-optimal solutions, which implies that the QoE for one user can not be improved without degrading other users' QoE.
Furthermore, the action space is an inter-dependent multi-task space.
This implies that tasks (\ie rate allocation and computation distribution) put constraints on each other (\eg the maximum achievable rate is constrained by the rate adaptation decision).
Another challenge arises from varying $360^\circ$ video characteristics, which means the computational requirements may differ from one video to another. 
Finally, VR users experience time-varying and dynamic wireless network conditions (e.g., due to blockage~\cite{ghazikor-2024-interference}, interference~\cite{Ghazikor-2023-Exploring}, etc.) that impact the streaming data rate. 
% While, mmWave networks enable multi-Gbps transmission rates by benefiting from high bandwidth and employing various technologies such as directional beamforming techniques, they suffer from higher path loss due to free space, oxygen absorption, and/or penetration losses, which introduces uncertainty into the system.
Thus, a decision-making algorithm that incorporates various dynamic and time-varying conditions in terms of video quality, wireless network, and available computational resources is desirable.
\section{Multi-Task Decision-Making Framework}\label{sec:solution}
\vspace{-.05in}
To tackle the challenges mentioned above, we present the MTRC framework for multi-task rate adaptation and computation distribution algorithm.
% is modeled by a deep neural network. 
The MTRC solution leverages a state-of-the-art DRL method to learn the optimal rate adaptation and computation distribution policy, thereby maximizing QoE for all users.
% Leveraging the state-of-the-art DRL method, the MTRC agent learns the optimal multi-task
% This framework comprises two main components: (i) A joint rate adaptation and computation distribution agent, which is modeled by a deep neural network, and (ii) a proportional resource allocation algorithm, which allocates decoding/rendering resources to users that have been distributed to ECU.
% Upon receiving the state information for each GoP, our MTRC agent makes a joint rate allocation (\ie number of enhancement layers to prepare) and computation distribution (\ie decode/render the GoP on ECU or headsets) decision.
Next, we describe the proposed framework.
% \vspace{-.05in}

\textbf{MTRC Agent:}
% \vspace{-.05in}
At each time step and after receiving the request for each GoP from all users,
the DRL agent observes the state $s^m$ of all users. The state input $s^m$ includes \emph{playback statistics}, and \emph{video information}.
% \begin{itemize}
    % \item \textbf{\emph{Playback Statistics:}}
    The video playback statistics include six critical metrics for each user to fully describe the $360^\circ$ video playback status.
These metrics are past throughput, past decoding time, past transmission time, past rendering time, last allocated rate $e^{m-1}_n$, and current buffer level $B^m_n$.
We take the future GoPs size and number of remaining GoPs for each user as video information.
This will help our MTRC model to distinguish between videos with different spatio-temporal characteristics.
All of these metrics are collected for all users and stacked together to represent the state information.

Once the state $s^m$ is observed, the agent chooses an action $a^m$ to decode, render, and transmit GoPs.
In this case, the MTRC agent takes a joint action $a^m=(e^m, \phi^m)$.
% combined of rate allocation action $r_m$ and computation distribution action $\phi_m$ for all the users.
The rate allocation action $e^m \in \{0, 1\}^{N\times L}$ determines how many enhancement layers will be streamed to each user. 
The computation distribution action $\phi^{m} \in \{0,1\}^{N\times 3}$ determines where each user decodes and/or renders the $m^{th}$ GoP. 
% For example, $\phi_{m,2}^n = 1$ determines that decoding and rendering happen at the VR headset, thereby freeing up some ECU computation resources for other users. 

% After taking the action,  the state of the environment changes, and the agent receives a reward vector $r_m$ that gives the reward for each user.
% The goal of our agent is to maximize the expected cumulative discounted reward for all users. 
% We employ the change in users' perceived QoE at each step
% caused by the last action performed by the MTRC agent 
% as the reward term. This is defined as $ r^{m+1}_n = QoE^{m+1}_n - QoE^m_n$. 
% This reward captures the changes in the perceived QoE as a result of the last action performed by the MTRC agent, which enables the agent to learn the actions that lead to improvement in QoE.

% \bb{
After taking the action,  the state of the environment changes, and the agent receives a reward vector $r^m$ that gives the reward for each user.
The goal of our agent is to maximize the expected cumulative discounted reward for all users.
However, the definition of the reward in such an unconstrained optimization problem is not straightforward.
Hence, we modify the optimization problem outlined in Eq. \ref{eq:qoe-maximization} in order to capture the constraint violation and penalize the MTRC agent for not meeting the constraints.
Leveraging the duality principle, we write the Lagrangian dual problem of Eq. \ref{eq:qoe-maximization}:
\begin{equation}
\label{eq:qoe-lagrange}
% \small
\begin{aligned}
    \min_{\mu_0, \mu_1} \; \max_{\mathbf{\phi_n, e_n}} & \;\;   \underbrace{Q(\mathbf{e_n})   
    + \mu_0 \left( \mathcal{H}_0 - S(\mathbf{e_n}, \mathbf{\phi_n}) \right) 
    + \mu_1 \left( \mathcal{H}_1 - V(\mathbf{e_n}) \right)}_{:= QoE_n^M}\\
    \text{s.t.} & \;\; \text{Eqs.} \;\; \ref{eq:buffert-b}, \ref{eq:buffert-t}, \ref{eq:comp-const},
\end{aligned}
\end{equation}
which is equal to:
\begin{equation}
\label{eq:qoe-lagrange-mu}
% \small
\begin{aligned}
    \min_{\mu_0, \mu_1} & \quad\;   Q(\mathbf{e_n^\ast})   
    + \mu_0 \left( \mathcal{H}_0 - S(\mathbf{e_n^\ast}, \mathbf{\phi_n^\ast}) \right) 
    + \mu_1 \left( \mathcal{H}_1 - V(\mathbf{e_n^\ast}) \right)\\
    \text{s.t.} & \quad \text{Eqs.} \;\; \ref{eq:buffert-b}, \ref{eq:buffert-t}, \ref{eq:comp-const},
\end{aligned}
\end{equation}
where $\mathbf{e_n^\ast}$ and $\mathbf{\phi_n^\ast}$ are optimal rate allocation and computation distribution actions for user $n$.
Then, the optimal coefficients for rebuffering time and quality variation, which correspond to Lagrangian multipliers, can be obtained by solving Eq. \ref{eq:qoe-lagrange}:
\begin{equation}
\label{eq:mu-solutions}
% \small
\begin{aligned}
    \mu_0^\ast &= \argmin_{\mu_0} \;\; \mu_0 \left( \mathcal{H}_0 - S(\mathbf{e_n^\ast}, \mathbf{\phi_n^\ast}) \right), \\
    \mu_1^\ast &= \argmin_{\mu_1} \;\; \mu_1 \left( \mathcal{H}_1 - V(\mathbf{e_n^\ast}) \right). 
\end{aligned}
\end{equation}
Here, by minimizing the loss functions presented in Eq. \ref{eq:mu-solutions}, $\mu_0$ and $\mu_1$ are updated to increase (or decrease) if the rebuffering time or quality variation is higher (or lower) than the target rebuffering time and quality variation $\mathcal{H}_0$ and $\mathcal{H}_1$, respectively.
This modification in the optimization problem effectively handles the reward magnitude change over time during training. 
Hence, one needs to set only the target rebuffering time $\mathcal{H}_0$ and the target quality variation $\mathcal{H}_1$ for each user and video, and then the coefficients $\mu_0$ and $\mu_1$ are automatically adjusted over time to meet the target constraints.

Now that the objective function in Eq. \ref{eq:qoe-lagrange} accounts for both rebuffering time and quality variation, we employ the change in users' perceived QoE at each step
% caused by the last action performed by the MTRC agent 
as the reward term.
This is defined as $ r^{m+1}_n = QoE^{m+1}_n - QoE^m_n$. 
% We employ Eq. \ref{eq:surrogate-qoe-maximization} as the reward function for the MTRC agent.
This reward captures the changes in the perceived QoE as a result of the last action taken by the MTRC agent, which enables the agent to learn the actions that lead to improvement in QoE.

\begin{figure}[t]
    \centering
    \includegraphics[width=.8\linewidth, trim={0cm 1.8cm 5.4cm 0cm},clip]{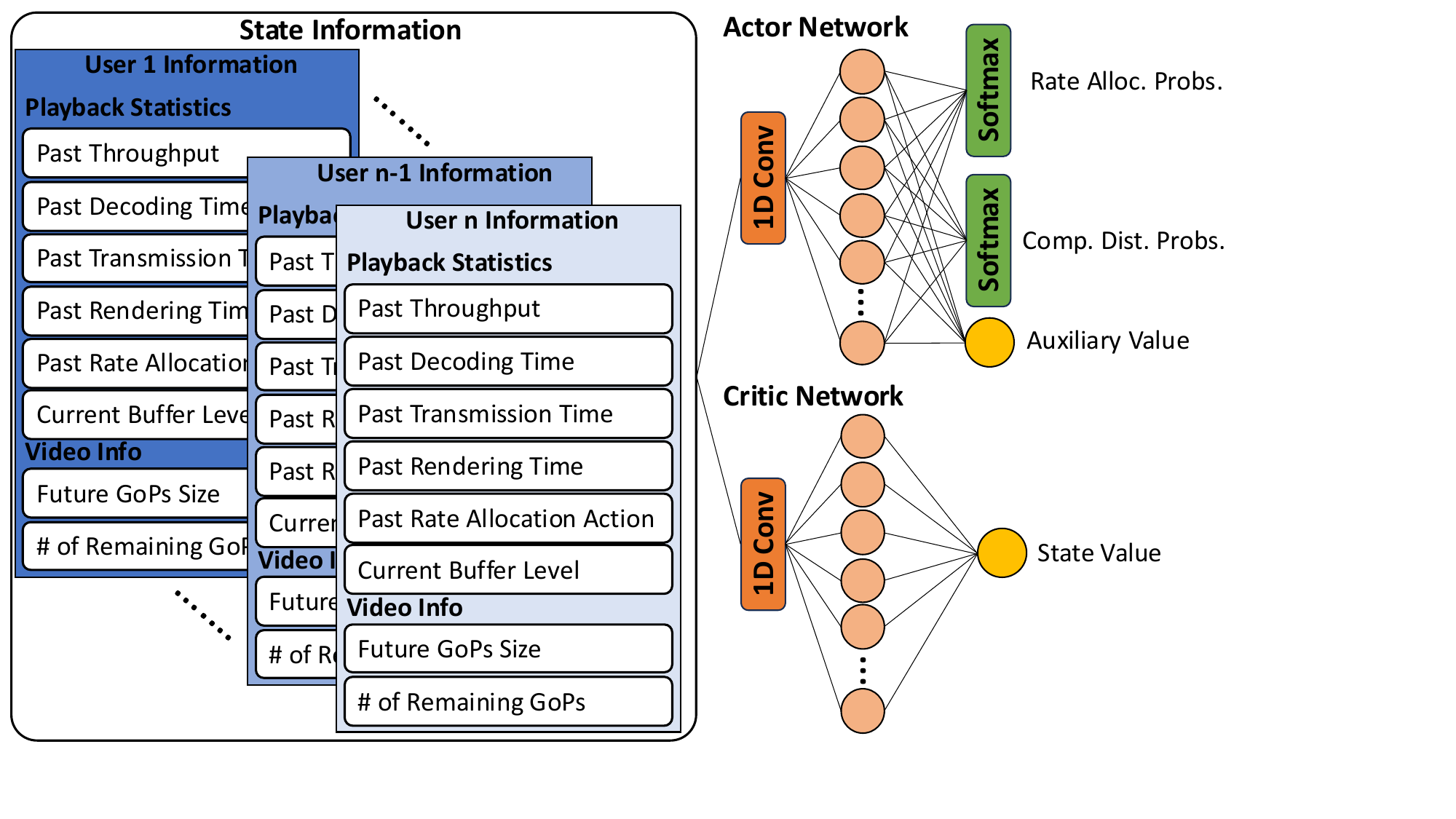}
    \vspace{-.05cm}
    \caption{MTRC Architecture}
    \label{fig:deep-vr-arch}
    \vspace{-.05in}
\end{figure}
% \vspace{-.6cm}
% \vspace{-.1in}
\textbf{DRL Implementation:}
% \vspace{-.1in}
We develop a sample-efficient multi-task DRL algorithm to tackle the complexity of the defined multi-task QoE maximization problem. 
The MTRC agent is composed of an actor network $\omega$ and a critic network $\omega_v$.
The actor network outputs the probabilities of both rate allocation action $\pi_{\omega}^{e}$ and computation distribution action $\pi_{\omega}^{\phi}$ for all users.
% , the probabilities of decoding action $\pi_{\omega}^{\psi}$, and the probabilities of rendering action $\pi_{\omega}^{\theta}$ 
% for all the users.
The actor network also outputs an auxiliary vector that estimates the state value for each user separately.
The critic network outputs the estimated state value for each user separately.
% This architecture reduces the interference between policy and value loss, while distilling features from the value function into the policy network \cite{cobbe-2020-ppg}.
% \textbf{Actor Network:}
% The actor network includes computation location actions for all users, resource allocation profiles, and a vector of $N$ scalar values named auxiliary value.
% The auxiliary value is used purely to train representations for the policy, and it will be optimized by the auxiliary loss during the auxiliary phase.
% \bb{BB: Go over the NN architecture here in detail.}
% \textbf{Critic Network:} 
% The critic network output a vector, which learns an estimate of the accumulated reward, earned by each user, that will be used to train the actor network.
% \bb{BB: Go over the NN architecture of critic network in detail.
% \textbf{Policy Optimization:}
% Our proposed solution is mainly inspired by \cite{cobbe-2020-ppg, Deheng-2020-Mastering, Tianchi-2023-Buffer}.
Inspired by \cite{Deheng-2020-Mastering, Tianchi-2023-Buffer}, we employ a two-phase training procedure, composed of a policy training phase and an auxiliary training phase~\cite{cobbe-2020-ppg}.
% This work, later, was adopted by \cite{Deheng-2020-Mastering} to tackle the problem of large action spaces, and buffer wastage in video streaming applications \cite{Tianchi-2023-Buffer}.
% Inspired by \cite{Deheng-2020-Mastering, Tianchi-2023-Buffer}, we adopt \cite{cobbe-2020-ppg} to train our MTRC agent.
% The training process is composed of policy training phase and value training phase.
% In the policy training phase, 
% The state of the art sample-efficient DRL algorithm in which the training process is composed of a policy training phase and an auxiliary training phase.
% The considered setting, in \cite{cobbe-2020-ppg}, is a single task setting, where the DRL agent learns to perform a single task in an environment.
% We employ \cite{Deheng-2020-Mastering} to train our multi-task QoE maximization MTRC Streamer.
% \cite{schulman-2017-proximal}
In the policy training phase, 
% we update both actor and critic networks.
the actor and critic networks are trained by Dual-Clip PPO \cite{schulman-2017-proximal}:
\begin{equation}
% \small
\label{eq:d-clip-loss}
    \begin{aligned}
        \mathcal{L}^{DClip} = \hat{\mathbb{E}}
            \left[ \indicator(\hat{A}^m < 0)\max(\mathcal{L}^{PPO}, c\hat{A}^m) + \indicator(\hat{A}^m \geq 0)\mathcal{L}^{PPO} \right].
    \end{aligned}
\end{equation}
Here, $\indicator()$ denotes a binary indicator function, and $\mathcal{L}^{PPO}$ represents the surrogate vanilla PPO loss
% function introduced in  \cite{schulman-2017-proximal} 
that is:
\begin{equation}
% \small
\label{eq:ppo-loss}
    \begin{aligned}
        \mathcal{L}^{PPO} = \mathcal{L}^{Clip}(\pi_{\omega}, \hat{A}^m) + \beta H_{\omega}(s^m) + \mathcal{L}^{Value},
    \end{aligned}
\end{equation}
where $H(s^m)$ is the entropy of all policies, and $\beta$ is the entropy weight, which jointly balance the tradeoff between exploration and exploitation during the learning process.
$\mathcal{L}^{Value}$ is the value network loss \cite{cobbe-2020-ppg}, and $\mathcal{L}^{Clip}$ is the Single-Clip policy loss: 
% Eq. \ref{eq:value-loss}.
\begin{equation}
% \small
\label{eq:clip-loss}
    \begin{aligned}
        \mathcal{L}^{Clip} &= \min \left[ \rho(\pi_{\omega}, \pi_{\omega_{old}}) \odot \hat{A}^m, clip(\rho(\pi_{\omega}, \pi_{\omega_{old}}), 1 \pm \varepsilon) \odot \hat{A}^m \right].
    \end{aligned}
\end{equation}
Here, $\hat{A}^m = r_m + \gamma V_{\omega_v}(s^{m+1}) - V_{\omega_v}(s^m)$ is the advantage function that is calculated based on the current state-value estimate and the discount factor $\gamma = 0.99$, 
and $\odot$ is the element-wise multiplication. $\rho(\pi_{\omega}, \pi_{\omega_{old}}) = \left[\rho_1, \rho_2, \dotsb, \rho_N \right]$ is a vector of size $N$ that measures the changes in the new policy w.r.t. the old policy for all users (\ie the joint probability ratio of the new policy to the old policy for the joint action):
% both computation distribution and rate allocation action:
% \ie the joint probability ratio of the new policy to the old policy for all the users:
% \begin{equation*}
% \small
%     \begin{aligned}
%          \rho(\pi_{\omega}, \pi_{\omega_{old}}) = \left[\rho^1, \rho^2, \dotsb, \rho^n \right].
%     \end{aligned}
% \end{equation*}
% Each element of this vector $\rho^i$ measures the changes for one user in the environment, \ie the joint probability ratio of the new policy to the old policy for both computation distribution and rate allocation action:
\begin{equation*}
% \small
    \begin{aligned}
        \rho_i =
        % \prod_{n=1}^{N}
        \frac{
            \pi_{\omega}^{\phi} (\phi^m_i|s^m)
        }{
            \pi_{\omega_{old}}^{\phi} (\phi^m_i|s^m)
        } 
        % \odot
        \frac{
            \pi_{\omega}^{e} (e^m_i|s^m)
        }{
            \pi_{\omega_{old}}^{e} (e^m_i|s^m)
        }.
    \end{aligned}
\end{equation*}
% $$
% \normalsize
% Note that both $\rho(.,.)$ and $\hat{A}^m$ are vectors of size $N$
In Eq. \ref{eq:clip-loss}, $\rho(\pi_{\omega}, \pi_{\omega_{old}}) \odot \hat{A}^m$ measures the gained/lost reward as the policy for each user changes, determining how a change in one user's policy affects the reward of other users.
% , \ie perceived QoE. 
% and $\rho(.,.)$ measures the change in the new policy w.r.t. the old policy, \ie the joint probability ratio of the new policy to the old policy for a multi-user setting:
% \small
% $$
% \begin{equation*}
% \small
%     \begin{aligned}
%         \rho(\pi_{\omega}, \pi_{\omega_{old}}) = \prod_{n=1}^{N}
%         \frac{
%             \pi_{\omega}^{\phi^n} (\phi^m_n|s^m)
%         }{
%             \pi_{\omega_{old}}^{\phi^n} (\phi^m_n|s^m)
%         }
%         % \frac{
%         %     \pi_{\omega}^{\psi^n} (\psi^m_n|s^m)
%         % }{
%         %     \pi_{\omega_{old}}^{\psi^n} (\psi^m_n|s^m)
%         % }
%         % \frac{
%         %     \pi_{\omega}^{\theta^n} (\theta^m_n|s^m)
%         % }{
%         %     \pi_{\omega_{old}}^{\theta^n} (\theta^m_n|s^m)
%         % }
%         .
%     \end{aligned}
% \end{equation*}
% $$
% \normalsize
In the auxiliary phase, we further optimize the actor and critic networks according to a joint objective function $\mathcal{L}^{Joint}$, which is composed of a behavioral cloning loss and an auxiliary value loss.
Due to space limitation, details of the auxiliary training phase are omitted.
Interested readers can refer to \cite{cobbe-2020-ppg}.

\begin{algorithm}[t]
\caption{MTRC Training Process}
\label{alg:training-alg}
% \textbf{Inputs:} \\
% \hspace*{\algorithmicindent}  $Q$ \;\,: A neural network initialized by $\theta_0$ \\
% \hspace*{\algorithmicindent}  $\mathcal{E}$ \,\,: Number of training epochs \\
% \hspace*{\algorithmicindent}  $\mathcal{B}$ \,: Task replay buffer \\
% \textbf{Algorithm:}
\begin{algorithmic}[1]
    \FOR{epoch $=1,2,...$}
        \STATE Perform rollout under current policy $\pi$
        % \STATE Compute value function target $V_{targ}$ for each state $s^m$
        \FOR{$i = 1,2,..., N_{Policy}$}
            \STATE Optimize Eq. \ref{eq:d-clip-loss} (\ie $\mathcal{L}^{DClip}$) w.r.t. $\omega, \omega_{v}$
        \ENDFOR
        \FOR{$i = 1,2,..., N_{aux}$}
            % \STATE Optimize Eq. \ref{eq:joint-loss} (\ie $\mathcal{L}^{Joint}$) wrt $\omega$
            % \STATE Optimize Eq. \ref{eq:value-loss} (\ie $\mathcal{L}^{Value}$) wrt $\omega_v$
            \STATE Optimize $\mathcal{L}^{Joint}$ w.r.t. $\omega$
            \STATE Optimize $\mathcal{L}^{Value}$ w.r.t. $\omega_v$
            % \STATE Optimize Eq. \ref{eq:value-loss} wrt $\omega_V$
            % \STATE Update  $\omega$ and $\omega_v$ according to auxiliary phase \cite{cobbe-2020-ppg}
        \ENDFOR
        \STATE Update $\mu_0$ and $\mu_1$ according to Eq. \ref{eq:mu-solutions}
    \ENDFOR
\end{algorithmic}
% \vspace{-.25in}
\end{algorithm}
Algorithm~\ref{alg:training-alg} presents the training process of MTRC agent, which continues for multiple iterations until convergence.
Each iteration is composed of four phases.
In the first phase, we perform the current policy $\pi_{\omega}$ on a randomized environment to collect new experiences (\ie rollout process).
We encapsulated our edge-assisted VR system into a gym-like environment, which allows the MTRC agent to interact with the system effectively.
The MTRC agent learns the policy through its interaction with the environment.
Both users' video and network conditions are randomized in this environment so that the agent learns the optimal computation distribution and rate adaptation policy for various videos and network conditions.
In the second phase, we update both the actor and critic networks.
We compute the Dual-Clip PPO loss $\mathcal{L}^{DClip}$, and use newly collected experiences (\ie resulted from the rollout process) to update the networks.
In the third phase, we use all collected experiences to update both the actor and critic networks by optimizing the behavioral cloning and value losses.
Finally, in the fourth phase, we update $\mu_0$ and $\mu_1$ according to Eq. \ref{eq:mu-solutions}.

% \begin{figure*}[t]
% % \vspace{-0.1cm}
% \centering
%     \begin{subfigure}[t]{0.3\linewidth}
%          \centering
%          \includegraphics[width=\linewidth]{figs/globe/training_qoe_mixed_2.pdf}
%          \vspace{-.7cm}
%          \caption{Average user QoE}
%          \label{fig:net-analysis}
%      \end{subfigure}
%      \begin{subfigure}[t]{0.3\linewidth}
%          \centering
%          \includegraphics[width=\linewidth]{figs/globe/training_psnr_mixed_2.pdf}
%               \vspace{-.7cm}
%          \caption{Average PSNR perceived by users}
%          \label{fig:net-analysis}
%      \end{subfigure}
%      \begin{subfigure}[t]{0.3\linewidth}
%          \centering
%          \includegraphics[width=\linewidth]{figs/globe/training_stall_time_mixed_2.pdf}
%                   \vspace{-.7cm}
%          \caption{Average rebuffering time}
%          \label{fig:res-alloc-analysis}
%      \end{subfigure}
%      % \vspace{-.2cm}
% \caption{The MTRC agent learns to maximize the QoE (\ie weighted average of PSNR, rebuffering time, and quality variation) in a VR arena with $N=6$ users over 2000 episodes. The MTRC agent outperforms two Pensieve variations due to its computation distribution ability.}
% \label{fig:training-performance}
% \vspace{-0.6cm}
% \end{figure*}

\vspace{-.05in}
\section{Evaluation}\label{sec:evaluation}
\vspace{-.05in}
\textbf{Simulation and Training:} In our simulations, we employ a full UHD $360^\circ$ video dataset \cite{Chakareski-2021-Full}.
This dataset includes 9 videos with various spatio-temporal characteristics.
Each video is represented using the multi-layer $360^\circ$ model presented in Section \ref{sec:system}, and the video frames are partitioned into a $8\times8$ grid.
% Table \ref{tab:video-info} presents the average and variance of the bitrate for the videos included in the dataset.
% All the videos in this dataset are 
% Bitrate information for multiple video quality per tile is provided, which enables us to 
% Bitrate information for multiple video quality of each video is available in
% Bitrate information for seven layers of increased immersion fidelity for each tile is provided.
% Moreover, head navigation data for multiple users is provided, which enables us to calculate each user's viewport location.
Bitrate information for seven layers, each offering progressively higher levels of immersion fidelity for each tile, is provided.
Additionally, head movement data for multiple users is included, allowing us to determine the viewport location for each user.
% which enables us to simulate the multi-user VR arena.
% We use this head navigation data to calculate each user's viewport location.
% We use the viewport information to determine which tiles are in the user's viewport, and calculate the video bitrate.
% \textbf{Network Trace Dataset:}
Moreover, we use a dataset of wireless network throughput traces~\cite{Narayanan-2021-Variegated}, which were collected from commercial operators (T-Mobile and Verizon) in two cities in the U.S. 

\textbf{Baselines:} We evaluate our proposed framework through an extensive simulation against Pensieve \cite{Mao-2017-Neural}, a state-of-the-art rate adaptation algorithm.
% two learning-based methods, \ie PPG and PPO, two static resource allocation mechanisms, \ie PropECU and Headset, and the optimal computation distribution mechanism, \ie Optimal.
% Two of the baselines are variations of the-state-of-the-art rate adaptation algorithm \cite{Mao-2017-Neural}, Pensieve.
Pensieve is designed for 2D video streaming applications, and only adjust the video rate based on user state, while our method is a multi-task rate adaptation and computation distribution algorithm. 
Thus, we employ two variants of Pensieve, namely ECU-Pensieve and Headset-Pensieve.
ECU-Pensieve performs all the computations (\ie decoding and rendering) on the ECU, while Headset-Pensieve performs all the computations on the users' headset.
Both use the original Pensieve with no modification to adaptively adjust the users' bitrates.
% The computational requirements for 2D videos are different from 3D videos, thus 
% since the decoding and computation can take place both on the ECU and users' headset.
% The first baseline is the-state-of-the-art adaptive rate adaptation algorithm \cite{Mao-2017-Neural} in various settings.
Moreover, we modify our MTRC framework and present two rate adaptation algorithms, ECU-R and Headset-R.
ECU-R and Headset-R use a neural network with the same architecture as shown in Fig. \ref{fig:deep-vr-arch} for rate adaptation, except that the computation distribution is not decided by the neural network and all computations are performed on the ECU or headsets, respectively.
We train our MTRC agent in an environment with $N=6$ VR users, ECU decoding speed of $Z^{dec.}_{ECU} = 7.5$~Gbps, ECU rendering speed of $Z^{rend.}_{ECU} = 20$ Gbps, headsets' decoding and rendering speeds of $Z^{dec.}_{n} = 0.2$ Gbps and $Z^{rend.}_{n} = 9.4$ Gbps, respectively.
The $360^\circ$ videos and network traces are randomly chosen in each episode of learning, and we train all the baselines for $5000$ episodes.
Then, we run $300$ testing episodes and report the performance of our MTRC method compared to other baselines in the testing stage.

\begin{figure}[t]
% \vspace{-0.1cm}
\centering
    \begin{subfigure}[t]{0.48\linewidth}
         \centering
         \includegraphics[width=\linewidth]{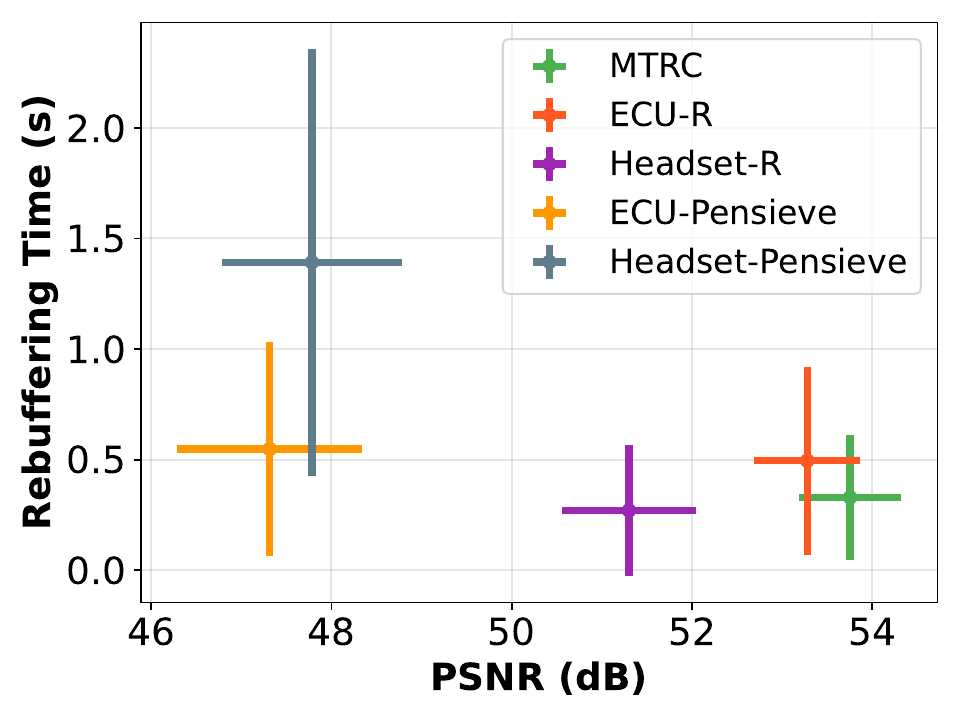}
         \vspace{-.7cm}
         \caption{RT vs PSNR}
         \label{fig:rt-vs-psnr}
     \end{subfigure}
     \begin{subfigure}[t]{0.48\linewidth}
         \centering
         \includegraphics[width=\linewidth]{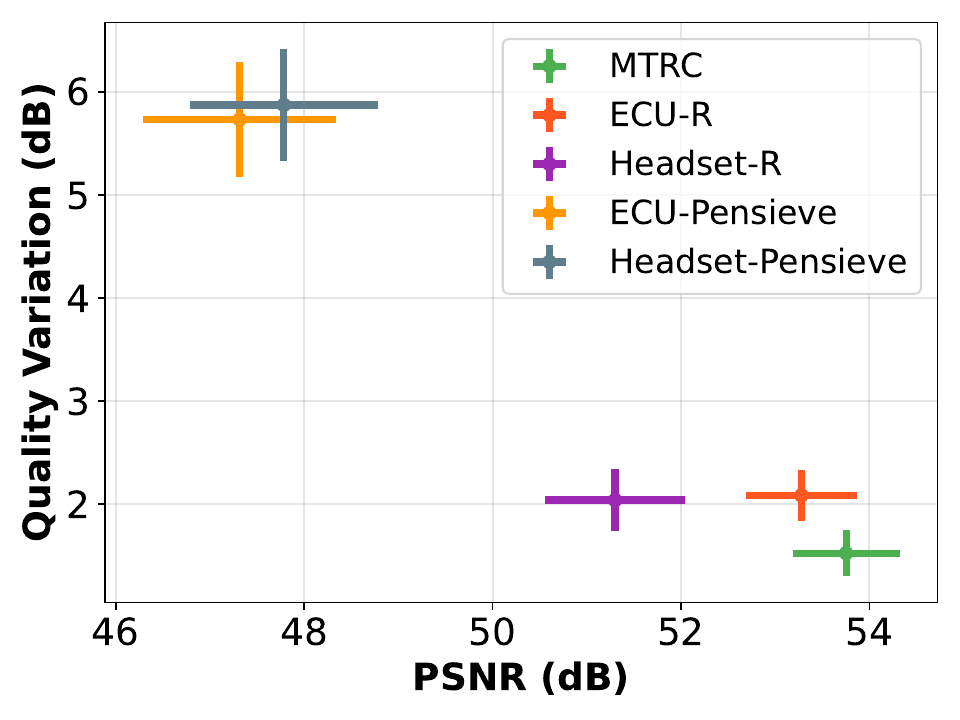}
              \vspace{-.7cm}
         \caption{QV vs PSNR}
         \label{fig:qv-vs-psnr}
     \end{subfigure}
     % \vspace{-.2cm}
\caption{Performance trade-offs between rebuffering time, quality variation, and PSNR during testing stage.
% The MTRC agent learns to maximize the QoE (\ie weighted average of PSNR, rebuffering time, and quality variation) in a VR arena with $N=6$ users over 2000 episodes. The MTRC agent outperforms two Pensieve variations due to its computation distribution ability.
}
\label{fig:testing-performance}
% \vspace{-0.6cm}
\end{figure}

\textbf{Testing Performance:}
% Fig. \ref{fig:testing-performance} and Table \ref{tab:performance-metrics} compare the performance of the MTRC agent to the baselines. 
Fig. \ref{fig:testing-performance} demonstrates the performance trade-offs between rebuffering time, quality variation, and PSNR.
Each point demonstrates the average rebuffering time (or quality variation) and PSNR experienced by the users. The vertical and horizontal bars represent the standard deviation of the rebuffering time (or quality variation) and PSNR, respectively.
A small rebuffering time (or quality variation) and high PSNR with small variation is desirable, which is represented by a point in the lower right corner of these plots. 
% Fig. \ref{fig:rt-vs-psnr} shows that the MTRC agent achieves the smallest rebuffering time and the highest PSNR.
% Fig. \ref{fig:qv-vs-psnr} reports the the average quality variation and PSNR experienced by users, and the vertical and horizontal bars represent the standard deviation of the quality variation and PSNR, respectively. Similary, a small quality variation and high PSNR with small variation is desirable, which is achieved by our MTRC agent.
Overall, the MTRC agent achieves an improvement of $5.97$ dB to $6.44$ dB in PSNR, $1.66$X to $4.23$X improvement in rebuffering time, and $4.21$ dB to $4.35$ dB improvement in quality variation.

Furthermore, Table \ref{tab:video-qoe} reports the average and standard deviation of PSNR, rebuffering time, and quality variations for groups of users that play a specific video. 
Table \ref{tab:video-qoe} shows that our proposed method is able to capture spatio-temporal characteristics of various $360^\circ$ videos and provide a fair QoE for groups of users with various requirements.

\begin{table}[htbp]
    \centering
    \resizebox{.95\linewidth}{!}{
    \begin{tabular}{|c|l|c|c|c|}
        \hline
        \textbf{\#} & \textbf{Video Name} & \textbf{PSNR [dB]} & \textbf{RT [Sec.]} & \textbf{AQV [dB]} \\
        \hline
        % 0 & Academic & $52.51 \pm 0.67$ & $0.18 \pm 0.39$ & $1.45 \pm 0.44$ \\
        % 1 & Basketball & $54.08 \pm 0.99$ & $0.32 \pm 0.49$ & $2.28 \pm 0.53$ \\
        % 2 & Bridge & $55.31 \pm 0.92$ & $0.63 \pm 0.98$ & $1.00 \pm 0.38$ \\
        % 3 & Gate Night & $55.61 \pm 0.89$ & $0.39 \pm 0.61$ & $1.18 \pm 0.40$ \\
        % 4 & Runner & $52.90 \pm 0.57$ & $0.23 \pm 0.41$ & $1.62 \pm 0.41$ \\
        % 5 & Siyuan & $53.42 \pm 0.65$ & $0.09 \pm 0.31$ & $1.70 \pm 0.37$ \\
        % 6 & South Gate & $53.81 \pm 1.18$ & $0.26 \pm 0.75$ & $1.76 \pm 0.52$ \\
        % 7 & Studyroom & $54.73 \pm 0.79$ & $0.17 \pm 0.35$ & $1.17 \pm 0.38$ \\
        % 8 & Sward & $51.41 \pm 0.98$ & $0.71 \pm 1.01$ & $1.53 \pm 0.75$ \\
        0 & Academic & $52.05 \pm 1.10$ & $0.13 \pm 0.32$ & $1.58 \pm 0.41$  \\
1 & Basketball & $53.72 \pm 1.34$ & $0.34 \pm 0.60$ & $2.29 \pm 0.44$  \\
2& Bridge & $55.00 \pm 0.85$ & $0.57 \pm 1.17$ & $1.27 \pm 0.36$  \\
3& Gate Night & $55.39 \pm 1.09$ & $0.25 \pm 0.62$ & $1.33 \pm 0.35$  \\
4& Runner & $52.56 \pm 0.78$ & $0.25 \pm 0.57$ & $1.81 \pm 0.40$  \\
5& Siyuan & $52.96 \pm 0.98$ & $0.10 \pm 0.26$ & $1.88 \pm 0.40$  \\
6& South Gate & $53.22 \pm 1.47$ & $0.26 \pm 0.65$ & $1.99 \pm 0.48$  \\
7& Studyroom & $54.23 \pm 1.06$ & $0.13 \pm 0.33$ & $1.36 \pm 0.38$  \\
8& Sward & $50.28 \pm 1.97$ & $0.68 \pm 1.02$ & $1.83 \pm 0.62$ \\
        \hline
    \end{tabular}
    }
    \caption{QoE percived by groups of users who play an specific video}
    \label{tab:video-qoe}
    % \vspace{-.1in}
\end{table}

\begin{table}[t]
    \centering
    \resizebox{.95\columnwidth}{!}{
    \begin{tabular}{|l|c|c|c|}
        \hline
        \textbf{Network Condition} & \multicolumn{3}{c|}{\textbf{Low Throughput: $662.22 \pm 359.25$ Mbps}} \\
        \hline
        \textbf{Baseline} & \textbf{PSNR [dB]} & \textbf{RT [s]} & \textbf{AQV [dB]} \\
        \hline
        MTRC & $\mathbf{53.63 \pm 1.51}$ & $0.38 \pm 0.74$ & $\mathbf{1.57 \pm 0.62}$ \\
        ECU-R & $53.21 \pm 1.42$ & $0.64 \pm 1.25$ & $2.12 \pm 0.59$ \\
        Headset-R & $51.29 \pm 2.19$ & $\mathbf{0.31 \pm 0.79}$ & $2.04 \pm 1.10$ \\
        ECU-Pensieve & $47.38 \pm 2.58$ & $0.65 \pm 1.17$ & $5.71 \pm 1.40$ \\
        Headset-Pensieve & $47.82 \pm 2.49$ & $1.51 \pm 2.51$ & $5.86 \pm 1.43$ \\
        \hline
        \textbf{Network Condition} & \multicolumn{3}{c|}{\textbf{High Throughput: $1454.55 \pm 381.04$ Mbps}} \\
        \hline
        % \textbf{Baseline} & \textbf{PSNR [dB]} & \textbf{RT [s]} & \textbf{AQV [dB]} \\
        % \hline
        MTRC & $\mathbf{53.96 \pm 1.51}$ & $0.06 \pm 0.05$ & $\mathbf{1.45 \pm 0.55}$ \\
        ECU-R & $53.71 \pm 1.55$ & $0.06 \pm 0.05$ & $1.90 \pm 0.56$ \\
        Headset-R & $51.52 \pm 1.95$ & $0.09 \pm 0.11$ & $2.00 \pm 1.15$ \\
        ECU-Pensieve & $47.22 \pm 2.69$ & $\mathbf{0.06 \pm 0.04}$ & $5.78 \pm 1.49$ \\
        Headset-Pensieve & $47.74 \pm 2.56$ & $0.83 \pm 1.67$ & $5.90 \pm 1.45$ \\
        \hline
    \end{tabular}
    }
    \caption{PSNR, rebuffering time, and quality variation for two groups of users in different network conditions}
    \label{tab:net-condition}
\end{table}

\textbf{Effect of Network Condition:}
Table \ref{tab:net-condition} demonstrates the effect of the network condition on QoE perceived by users.
To generate this table, we analyzed the QoE perceived by two groups of users in our simulation.
% To generate this table, we grouped users into two groups based on their network condition.
The first group experiences a network condition that leads to low throughput with an average and standard deviation of $662.22\pm359.25$ Mbps, while the second group experiences high throughput with an average and standard deviation of $1454.55\pm381.04$ Mbps.
% The first group experiences an average throughput of $662.22$ Mbps, while the
We report the average and standard deviation of the PSNR (\ie video quality), rebuffering time, and quality variation for these two groups of users.
Overall, the first group experiences an improvement of $5.81$ dB to $6.25$ dB in PSNR, $1.71$X to $3.97$X improvement in rebuffering time, and $4.14$ dB to $4.29$ dB improvement in quality variation.
The second group experiences even more improvements. The MTRC agent demonstrates an improvement of $6.22$ dB to $6.74$ dB in PSNR, a up to $13.83$X improvement in rebuffering time, and a $4.33$ dB to $4.45$ dB improvement in quality variation for this group.

\vspace{-.05in}
\section{Conclusion}\label{sec:conclusion}
\vspace{-.05in}
In this paper, we considered the problem of multi-task rate adaptation and computation distribution in a VR arena for a $360^\circ$ video streaming platform, where a learning-based multi-task agent decides on the video bitrate allocated to each user and computation distribution (\ie whether each video segment should be decoded/rendered on the ECU or on the headset). 
% The ECU decides between performing users' computation on ECU itself or on the users' headsets.
The overall objective is to maximize the QoE of users under dynamic and time-varying conditions in terms of video requests, available computational resources, and communication bandwidth. 
% However, there are multiple challenges in solving this problem, such as stochasticity of the system due to dynamic spatio-temporal characteristics of $360^\circ$ videos and varying network condition; interdependency of actions \eg the allocated rate may limit the computation distribution action and vice versa; and multi-objectivity of the problem which means the QoE for one user can not be improved without degrading other users' QoE.
Using the state-of-the-art DRL algorithm, we developed MTRC that utilizes playback statistics and video information to make a joint rate adaptation and computation distribution decision. 
% Our MTRC agent, located in the ECU, decides whether each GoP needs to be decoded and/or rendered on the ECU or on the user's headset.
Through numerical simulation using real-world network traces and $360^\circ$ video information, we showed that the MTRC agent learns to balance the existing trade-offs in the system and outperforms the state-of-the-art rate adaptation algorithm. 
Specifically, MTRC demonstrates an improvement of $5.97$ dB to $6.44$ dB in PSNR, $1.66$X to $4.23$X improvement in rebuffering time, and $4.21$ dB to $4.35$ dB in quality variation.  
% Our current solution relies on the proportional computation resource allocation on the ECU. In the future, we will optimally allocate these resources to users. 
% \jacob{We should try to highlight here our performance advances over the state-of-the-art. The conclusion can be shortened.}
% to distribute the computation among users' headset and ECU to achieve a close-to-optimal performance.
% Specifically, the MTRC improves the users' perceived video quality by $3.08 dB$ to $4.49 dB$
% 
% outperforms other learning-based and static computation distribution algorithms.
% provides an average of $1.9\%$ improvement in QoE perceived by users per GoP compared to its rival baselines.
% superior QoE compared to its rival baselines, which leads to  
% \vspace{-.05in}
\section{Acknowledgement}
% \vspace{-.05in}
The material is based upon work supported in part by the National Science Foundation (NSF) grants 1955561, 2212565, and 2323189. Any opinions, findings, and conclusions or recommendations expressed in this material are those of the author(s) and do not necessarily reflect the views of NSF.
The work of Jacob Chakareski has been supported in part by NSF under awards CCF-2031881, ECCS-2032387, CNS-2040088, CNS-2032033, and CNS-2106150; by the National Institutes of Health (NIH) under award R01EY030470; and by the Panasonic Chair of Sustainability at the New Jersey Institute for Technology.

% \vspace{-.3cm}
% \Urlmuskip=0mu plus 1mu\relax
\small
{
\bibliographystyle{IEEEtranN}
\bibliography{main}
}
\end{document}